\documentclass[cameraready]{Interspeech}

\usepackage{spconf,amsmath,graphicx,hyperref}
\usepackage[table]{xcolor}
\usepackage{amsmath,amsfonts}
\usepackage{algorithm}
\usepackage{algorithmicx}
\usepackage{algpseudocode}
\usepackage{array}
\usepackage[caption=false,font=normalsize,labelfont=sf,textfont=sf]{subfig}
\usepackage{textcomp}
\usepackage{stfloats}
\usepackage{url}
\usepackage{xcolor}
\usepackage{verbatim}
\usepackage{graphicx}
\usepackage{cite}
\usepackage{verbatim}
\usepackage{multirow}
\usepackage{makecell}
\usepackage{booktabs}
\usepackage{threeparttable}
\usepackage{amssymb}
\usepackage{bbding}
\usepackage{listings}
\usepackage{float}
\usepackage{changepage,subfig}
\usepackage{amssymb} % 
\usepackage{pifont} % 在导言区添加
\newcommand{\cmark}{\ding{51}}% 对勾
\newcommand{\xmark}{\ding{55}}% 叉号
\makeatletter
\title{Spatially-Augmented Sequence-to-Sequence Neural Diarization for Meetings}

\author[affiliation={1}, orcid=0009-0005-0892-5301]{Li}{Li}
\author[affiliation={3}, orcid=0000-0002-4733-3596]{Ming}{Cheng}
\author[affiliation={1}]{Juan}{Liu}
\author[affiliation={1,2}, orcid=0000-0002-4733-3596, correspondingauthor]{Ming}{Li}

\address{
    $^1$ School of Artifcial Intelligence, Wuhan University, Wuhan, China \\
    $^2$ School of Artificial Intelligence, The Chinese University of Hong Kong, Shenzhen, China \\
    $^3$ School of Computer Science, Wuhan University, Wuhan, China
}

\email{lili\_a0@163.com, ming.cheng@whu.edu.cn, liujuan@whu.edu.cn, ming.li.cuhksz@gmail.com}

\usepackage{comment}

\makeatletter
\def\name#1{\gdef\@name{#1\\}}
\name{\authorlist}

\def\maketitle{\par
    \begingroup
    \if@twocolumn
        \twocolumn[\@maketitle]
    \else 
        \newpage
        \global\@topnum\z@ 
        \@maketitle 
    \fi
    \@thanks
    \endgroup
    \ifcameraready
        \ifequalcontribution
            \def\thefootnote{*}
            \footnotetext{These authors contributed equally.}
        \fi
        \ifcorrespondingauthor
            \def\thefootnote{**}
            \footnotetext{Indicates the corresponding author.}
        \fi
    \fi
}

\def\@maketitle{%
    \newpage
    \newbox\titlebox
    \setbox\titlebox=\vbox{%
        \hsize=\textwidth
        \begin{center}
            {\large \bf \@title \par}
            \vskip 14pt
            {%
                \large
                \textit{\@name}%
                \vskip 11pt
                \@address
            }%
            \vskip 3pt
            {\small\texttt{\@email}}% 显示邮箱
            \par
        \end{center}
        \par
        \thispagestyle{empty}
        \vskip 9pt
    }
    
    \newdimen\titleboxheight
    \titleboxheight=\ht\titlebox
    \ifdim\titleboxheight<\defaulttitleboxheight
      \titleboxheight=\defaulttitleboxheight
    \fi
    
    \vbox to \titleboxheight{%
        \unvbox\titlebox
        \vfill
    }%
}
\makeatother

\begin{document}

\maketitle

\begin{abstract}

This paper proposes a Spatially-Augmented Sequence-to-Sequence Neural Diarization (SA-S2SND) framework, which integrates direction-of-arrival (DOA) cues estimated by SRP-DNN into the S2SND backbone. A two-stage training strategy is adopted: the model is first trained with single-channel audio and DOA features, and then further optimized with multi-channel inputs under DOA guidance. In addition, a simulated DOA generation scheme is introduced to alleviate dependence on matched multi-channel corpora. On the AliMeeting dataset, SA-S2SND consistently outperforms the S2SND baseline, achieving a 7.4\% relative DER reduction in the offline mode and over 19\% improvement when combined with channel attention. These results demonstrate that spatial cues are highly complementary to cross-channel modeling, yielding good performance in both online and offline settings.

\end{abstract}
\section{Introduction}
\label{sec:intro}

Speaker diarization aims to answer the ``who-spoke-when'' question~\cite{park2022review}, serving as a fundamental pre-processing step for downstream tasks (e.g., speech recognition)~\cite{kanda2020joint}. Despite notable progress, speaker diarization in meetings remains challenging due to overlapping speech, unreliable speaker embeddings, reverberation, etc.

Early studies on speaker diarization focus on modularized pipelines~\cite{wang2018speaker,landini2022bayesian}, which segment audio and cluster them by speaker embedding similarity. Assuming each segment contains only one speaker, these methods perform poorly on overlaps. End-to-End Neural Diarization (EEND) addresses this by formulating diarization as a multi-label prediction task, achieving greater robustness~\cite{fujita2019end_1, horiguchi2022encoder}. More recently, Target-Speaker Voice Activity Detection (TSVAD) combines modular and neural methods with strong performance~\cite{medennikov2020target, cheng2023target,wang2022similarity,wang2021dku}, while Sequence-to-Sequence Neural Diarization (S2SND) further advances online diarization~\cite{cheng2024sequence}. However, most approaches still rely solely on acoustic embeddings, which are often unreliable in real meetings. In contrast, spatial cues provide an orthogonal source of information, since speakers typically occupy different physical locations. This raises a key question: how can spatial cues from multi-channel recordings improve diarization?

In multi-channel processing, spatial cues can be integrated in three ways: 1) speech enhancement such as beamforming to generate cleaner inputs~\cite{Anguera,Boeddeker}, though they may often introduce distortions that harm speaker discrimination; 2) channel-fusion or attention modules that aggregate signals~\cite{Horiguchi2022}, but typically perform blind fusion rather than true localization; and 3) explicit features like direction-of-arrival (DOA) estimates~\cite{doa2008}. Among them, explicit DOA provides direct directional evidence to separate simultaneous speakers and thus holds greater promise for meeting diarization.

To effectively integrate DOA cues into diarization, two challenges must be solved: robust extraction of high-quality DOA under noise, reverberation, and multi-source conditions, and effective fusion of these cues into diarization models for both online and offline scenarios. Traditional localization methods (e.g., statistical filtering~\cite{doa2008}, HMM-based clustering~\cite{doa2021}, steered response power (SRP)~\cite{srp}) have been studied, but recent deep learning approaches show greater potential~\cite{nguyen2020robust,9413923,9746624}. In particular, SRP-DNN~\cite{9746624} learns direct-path phase differences and builds an enhanced SRP spectrum with iterative peak detection, achieving robust multi-speaker DOA estimation in adverse conditions. We therefore adopt SRP-DNN for DOA extraction and S2SND~\cite{cheng2024sequence} as the diarization backbone.

Building on these observations, we propose SA-S2SND (Spatially-Augmented S2SND). It injects DOA cues from SRP-DNN as explicit auxiliary inputs into the S2SND backbone, enhancing discriminability and improving performance in both online and offline modes. To improve generalization and decouple spatial cues from specific arrays, we introduce a simulated DOA strategy: real multi-channel speech is paired with estimated DOA, while simulated multi-channel speech (from single-channel data) is paired with simulated DOA. This reduces reliance on large-scale multi-channel corpora and improves adaptability to diverse arrays and conditions. Our main contributions are as follows:
\begin{itemize}
    \item We propose SA-S2SND, which integrates DNN-derived DOA as explicit spatial input to S2SND for online and offline diarization.
    \item We design a simulated-DOA method that decouples spatial cues from array design, enabling effective use of spatial information without large multi-channel corpora.
    \item We validate SA-S2SND on the AliMeeting dataset, showing consistent DER improvements over S2SND baselines in both modes.
\end{itemize}

\begin{figure*}[t]
\centering
  \includegraphics[width=\linewidth]{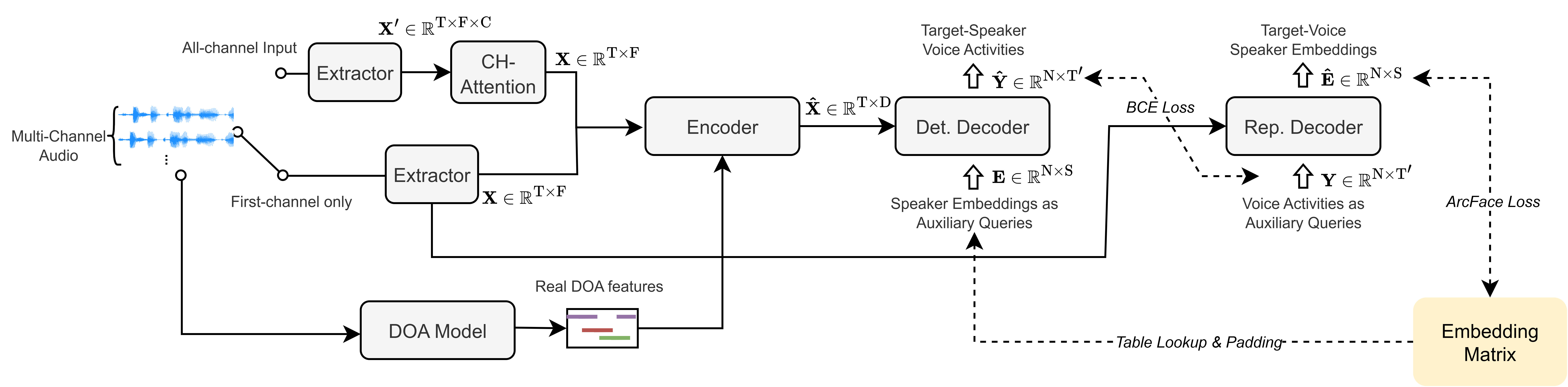}
  \vspace{-20pt}
  \caption{Spatially-Augmented Sequence-to-Sequence Neural Diarization (SA-S2SND) framework. \textit{Det.} and \textit{Rep.} denote the abbreviations of detection and representation, respectively.}
  \label{fig:framework}
  \vspace{-15pt}
\end{figure*}

\section{Methods}
\label{sec:method}

In this section, we introduce the method with two components: (1) revisiting direction-of-arrival (DOA) estimation from multi-channel inputs using SRP-DNN~\cite{9746624}; and (2) integrating these DOA cues into the sequence-to-sequence neural diarization (S2SND) backbone~\cite{cheng2024sequence} to form our proposed SA-S2SND framework.

\vspace{-0.2cm}
\subsection{DOA Estimation with SRP-DNN}

The first component is the direction-of-arrival (DOA) estimation module. We adopt SRP-DNN~\cite{9746624}, which estimates robust DOAs for multiple sources by learning direct-path inter-channel phase differences (DP-IPDs) with a causal CRNN. This model is lightweight (0.86M) and easy to retrain for different array configurations. For the $k$-th source, DOA is represented as 
$\boldsymbol{\theta}_{k}=\left[\theta_{k}^{\mathrm{ele}}, \theta_{k}^{\mathrm{azi}}\right]^{T}$, 
with elevation $\theta_{k}^{\mathrm{ele}}\in[0,\pi]$ and azimuth $\theta_{k}^{\mathrm{azi}}\in[-\pi,\pi)$.

Specifically, given $M$ microphones, the network extracts multi-channel short-time features (log-magnitude and phase of microphone pairs) and outputs a \emph{summed DP-IPD} vector $\widehat{R}_{mm'}(n)$ for each frame $n$. 
The training target is a weighted sum of direct-path IPD vectors:
\begin{equation}
R_{mm'}(n) = \sum_{k=1}^{K} \beta_k(n)\, r_{mm'}(\boldsymbol{\theta}_k(n)),
\label{eq:target}
\end{equation}
where $r_{mm'}(\boldsymbol{\theta})$ is the DP-IPD vector of candidate DOA and $\beta_k(n)\in[0,1]$ is the activity probability of source $k$. 
This avoids permutation issues, and the network minimizes the MSE between $\widehat R_{mm'}(n)$ (prediction) and $R_{mm'}(n)$. From $\{\widehat R_{mm'}(n)\}$, an SRP-style spatial spectrum is constructed:
\begin{equation}
P'(\boldsymbol{\theta};n)=\frac{2}{M(M-1)F}\sum_{m=1}^{M-1}\sum_{m'=m+1}^{M}
\Re\!\{\widehat R_{mm'}(n)^\mathsf{H} r_{mm'}(\boldsymbol{\theta})\},
\label{eq:srp_dnn_spec}
\end{equation}
where $F$ is the number of frequency bins.

To localize multiple sources, SRP-DNN applies an iterative detection-and-removal (IDL) strategy for each frame $n$: 1) obtain $\widehat R_{mm'}(n)$ via the CRNN; 2) construct $P'(\boldsymbol{\theta};n)$ and find the dominant peak as candidate DOA; 3) estimate its weight, namely the energy ratio of $\widehat R_{mm'}(n)$; 4) remove the source’s contribution from $\widehat R_{mm'}(n)$ and repeat until the predefined threshold is reached. This yields cleaner spatial spectra and more reliable DOA estimates under reverberant, noisy, and multi-speaker conditions, providing explicit spatial cues for the diarization backbone.

\subsection{Proposed SA-S2SND}

In this work, we propose SA-S2SND, extending our sequence-to-sequence diarization (S2SND)~\cite{cheng2024sequence} and its multi-channel variant (MC-S2SND)~\cite{cheng25b_interspeech} with cross-channel attention. Although these models support both online and offline diarization, they lack explicit spatial information that essential for separating simultaneous speakers in meetings. SA-S2SND addresses this by injecting DOA cues from SRP-DNN into the backbone.

\subsubsection{Architecture}

The SA-S2SND architecture is shown in Fig.~\ref{fig:framework}. Its backbone follows S2SND~\cite{cheng2024sequence}, which formulates diarization as a sequence-to-sequence task of joint \emph{speaker detection} and \emph{speaker representation}, consisting of four modules: an extractor, an encoder, and two coupled decoders.

The extractor is a ResNet~\cite{he2016deep} with segmental statistical pooling (SSP)~\cite{wang2022similarity}, producing frame-level embeddings. A Conformer encoder~\cite{gulati2020conformer} models long-range dependencies and outputs contextualized features \(\hat{X}\). On top, two symmetric decoders operate: 1) the representation decoder (Rep.) uses extractor outputs and voice-activity queries to generate target embeddings \(\widehat{E}\); 2) the detection decoder (Det.) uses encoder features \(\hat{X}\) and embedding queries to predict activities \(\widehat{Y}\). These decoders form an inverse pair, conditioning embeddings on activities and vice versa, thus avoiding clustering or PIT assignment.

Building on this backbone, SA-S2SND incorporates spatial cues. For each block, the DOA model outputs azimuths of active speakers, \(\theta_{k}^{\mathrm{azi}} \in \mathbb{R}^{{T''} \times \hat{N}}\) with \(\hat{N} \leq 2\). Since SRP-DNN produces azimuths in \([-180^\circ,180^\circ)\) at \(5^\circ\) resolution, we construct a DOA matrix \(\mathbf{O} \in \mathbb{R}^{{T''} \times A}\), where each row encodes the probability of activity at a given azimuth, aligned with \(\beta\) in Eq.~\eqref{eq:target}. Because SRP-DNN runs at coarser resolution (\(T''\)) than frame-level embeddings (\(T\)), \(\mathbf{O}\) is upsampled by nearest-neighbor interpolation, projected to dimension \(D\), and fused with acoustic embeddings via residual addition:
\begin{align}
    \mathbf{X} = \mathbf{X} + 
    \mathrm{Linear}_{\mathbb{R}^{A} \rightarrow \mathbb{R}^D}(\texttt{interpolate}(\mathbf{O})) / \sqrt{D},
\end{align}
where \(D\) is the hidden dimension. This enriches encoder representations with directional priors, analogous to positional encoding, enabling the model to distinguish speakers that overlap temporally but originate from different spatial locations.
 
In the multi-channel case, a cross-channel attention module is applied to extractor outputs, following MC-S2SND~\cite{cheng25b_interspeech}. While MC-S2SND aggregates frame-level speaker embeddings from different channels without preserving explicit spatial information. In contrast, SA-S2SND injects DOA cues as directional information, providing complementary spatial evidence and further improving performance.

\subsubsection{Training Process}

We divide training into two parts: \emph{Part A} trains the model with single-channel audio assisted by multi-channel DOA features; \emph{Part B} upgrades it to multi-channel audio (via cross-channel attention) with multi-channel DOA features. The model is trained on fixed-length blocks. Two loss terms are used jointly:
\begin{equation}
\mathcal{L}_{\mathrm{BCE}} = -\frac{1}{N T'}\sum_{n,t'} \big[y_{n,t'}\log\hat{y}_{n,t'} + (1-y_{n,t'})\log(1-\hat{y}_{n,t'})\big]
\label{eq:bce}
\end{equation}
\begin{equation}
\mathcal{L}_{\mathrm{Arc}} = \frac{1}{N}\sum_{n}-\log\frac{e^{s\cos(\theta_n+m)}}{e^{s\cos(\theta_n+m)}+\sum_{i\neq S_n} e^{s\cos\theta_i}}
\label{eq:arc}
\end{equation}
and the total loss is \(\mathcal{L}=\mathcal{L}_{\mathrm{BCE}}+\mathcal{L}_{\mathrm{Arc}}\).
A learnable embedding matrix \(E_{\mathrm{all}}\) and a non-speech embedding are used to form input enrollment queries via table lookup; absent slots are padded accordingly during mini-batching.

\textbf{Part A}: Single-channel model + Multi-channel DOA (Stages 1–3). The model takes single-channel audio, while DOA features are provided in parallel: SRP-DNN estimates DOA for real recordings, and pseudo-DOA is generated online for simulated mixtures using VAD with random DOA assignment and perturbation.

\begin{itemize}
\setlength{\itemsep}{0pt}
    \item \textbf{Stage 1.} 
    Initialize a ResNet extractor pretrained on speaker verification, freeze it, and train encoder/decoders on simulated mixtures with pseudo-DOA (LR=\textit{1e-4}).
    \item \textbf{Stage 2.} 
    Unfreeze the extractor and train on 80\% simulated (pseudo-DOA) + 20\% real (SRP-DNN DOA), jointly optimizing acoustic and spatial cues.
    \item \textbf{Stage 3.} 
    Fine-tune the entire single-channel SA-S2SND with a reduced LR of \textit{1e-5}.
\end{itemize}

\textbf{Part B}: Multi-channel model + Multi-channel DOA (Stages 4–5). 
The model is upgraded with cross-channel attention to process multi-channel audio, while DOA is still provided by SRP-DNN.

\begin{itemize}
\setlength{\itemsep}{0pt}
    \item \textbf{Stage 4.} 
    Add cross-channel attention on extractor outputs; freeze prior parameters and train only this branch. (LR=\textit{1e-4}).
    \item \textbf{Stage 5.} 
    Unfreeze all modules and jointly fine-tune the full multi-channel SA-S2SND with DOA (LR=\textit{1e-5}).
\end{itemize}

This two-part schedule clarifies the scope: Stages 1–3 train a single-channel model with DOA support, while Stages 4–5 extend it to multi-channel with injected DOA, ensuring a consistent path from acoustic-only to spatially-augmented modeling.

\subsubsection{Inference Process}

The inference of SA-S2SND follows the block-wise sliding-window scheme of S2SND~\cite{cheng2024sequence}. Each block has left, chunk, and right contexts with lengths $L_{\text{left}}$, $L_{\text{chunk}}$, and $L_{\text{right}}$, giving $L=L_{\text{left}}+L_{\text{chunk}}+L_{\text{right}}$. Setting chunk shift to $L_{\text{chunk}}$ yields latency $L_{\text{chunk}}+L_{\text{right}}$.

On top of the block-wise process, SA-S2SND incorporates DOA features: SRP-DNN azimuth cues are interpolated and fused with encoder outputs to align spatial and temporal information. The detection decoder takes a fixed embedding matrix 
\(E=[e_{\text{pse}},E_{\text{buf}},E_{\text{non}}]\) 
(pseudo-speaker, buffered speakers, and non-speech paddings), and outputs 
\(\hat{Y}=[\hat{y}_{\text{pse}},\hat{Y}_{\text{buf}},\hat{Y}_{\text{non}}]\). 
The representation decoder extracts updated embeddings 
\(\hat{E}=[\hat{e}_{\text{pse}},\hat{E}_{\text{buf}},\hat{E}_{\text{non}}]\). 
Buffer management with quality weighting accumulates reliable embeddings and registers new speakers once a threshold is exceeded.

After the first (online) pass, the final embedding buffer enables a second-pass offline decoding, where acoustic and DOA-enhanced representations jointly improve offline results. The inference follows S2SND~\cite{cheng2024sequence}, with DOA injection as the only modification.

\section{Experimental Setup}

\subsection{Datasets}

For simulated data, we use VoxCeleb2~\cite{Chung18b} (1M utterances, 6,112 speakers) with on-the-fly mixture generation. Real data come from AliMeeting~\cite{alimeeting} (104.75h train, 4h eval, 10h test) containing 8-channel far-field array and headset recordings. We use the far-field array signals after dereverberation with NARA-WPE~\footnote{\url{https://github.com/fgnt/nara_wpe}}. Additionally, we also use a compound dataset for real data: Alimeeting~\cite{alimeeting}, DIHARD III~\cite{ryant2020third}, VoxConverse~\cite{voxconverse}, MISP2022~\cite{chen2022misp}.
The model is trained on the training set, validated on the evaluation set, and tested without oracle VAD or collar tolerance.

\begin{figure}[t]
  \centering
  \subfloat[Simulated data\label{fig:simu}]{
    \includegraphics[width=0.95\linewidth]{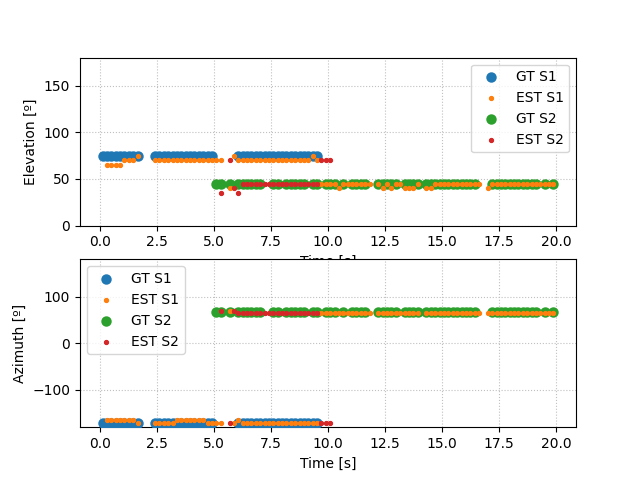}
  }\\
  \vspace{-10pt}
  \subfloat[Real data\label{fig:real}]{
    \includegraphics[width=0.95\linewidth]{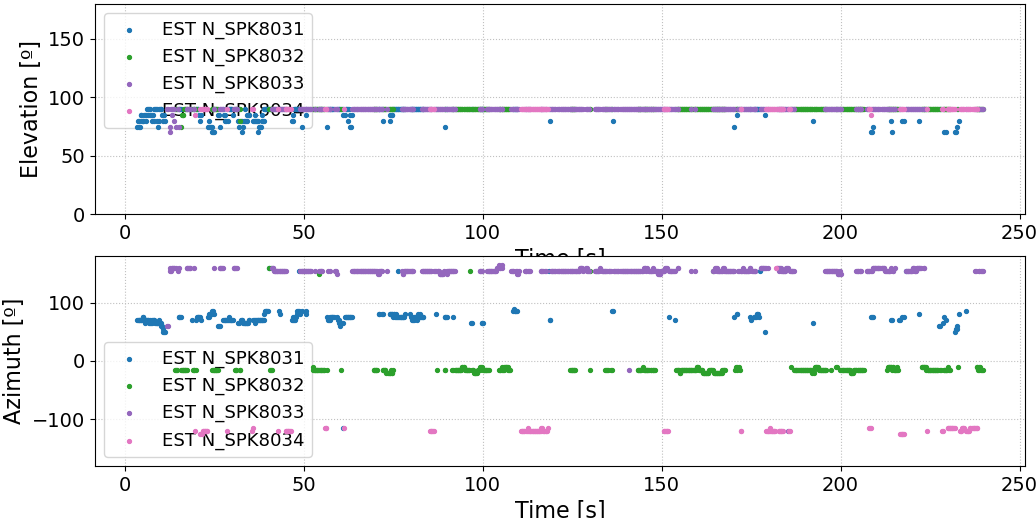}
  }
  \vspace{-0.2cm}
  \caption{Illustration of DOA estimation results for (a) simulated data, (b) real Alimeeting data.}
  \label{fig:both}
\vspace{-0.5cm}
\end{figure}

\vspace{-0.1cm}
\begin{table}[t]
\centering
\setlength{\tabcolsep}{25pt}
\caption{Overlap ratio of more than two speakers across datasets.}
\label{exp:ratio}
\begin{tabular}{lc}
\hline
\textbf{Dataset} & \textbf{Ratio (\%)} \\
\hline
AliMeeting & 6.0 \\
AISHELL-4 & 0.5 \\
AMI & 3.0 \\
\hline
\end{tabular}
\vspace{-0.5cm}
\end{table}

\begin{table*}[t]
	\centering
	\setlength{\tabcolsep}{4pt}
	\renewcommand{\arraystretch}{0.5}
	\caption{Performance on Alimeeting test sets with various model size and inferring conditions. The Diarization Error Rates (DERs) are reported without Oracle VAD and collar tolerance. Ch. Num. means the number of channels for S2SND.}
	%\vspace{-0.2cm}
	\label{exp:ali}
	\begin{threeparttable}[b]
	\begin{tabular}{lllllrrrrrr}
		\toprule
		\multirow{3}{*}{\textbf{ID}}
		& \multirow{3}{*}{\textbf{\makecell[l]{Model\\Size}}}
        & \multirow{3}{*}{\textbf{\makecell[l]{Ch. \\ Num.}}}
        & \multirow{3}{*}{\textbf{\makecell[l]{Usage \\of DOA}}}
		& \multirow{3}{*}{\textbf{\makecell[l]{Methods}}}
		& \multicolumn{2}{c}{\textbf{1-2 SPKs}}& \multicolumn{2}{c}{\textbf{2+ SPKs}}& \multicolumn{2}{c}{\textbf{Total}} \\
		\cmidrule(l{2pt}r{2pt}){6-7} \cmidrule(l{2pt}r{2pt}){8-9} \cmidrule(l{2pt}r{2pt}){10-11}
		&&&& &\textbf{\makecell[l]{Online\\DER (\%)}} & \textbf{\makecell[l]{Offline\\DER (\%)}} 
        & \textbf{\makecell[l]{Online\\DER (\%)}} & \textbf{\makecell[l]{Offline\\DER (\%)}} 
        & \textbf{\makecell[l]{Online\\DER (\%)}} & \textbf{\makecell[l]{Offline\\DER (\%)}} \\
        \midrule
		E1 & {Small(16.56M)} & {1} & \xmark & {S2SND} & 6.41 & 5.41 & 20.72 & 17.58 & 16.03 & 13.59 \\
	    \midrule
		E2 & {Small(16.56M+0.86M)} & {1} & \cmark & {SA-S2SND} & 6.35 & 5.40 & 19.75 & 16.10 & 15.35 & 12.59 \\
	    \midrule
		E3 & {Small(18M)} & {8} & \xmark & {S2SND} & \cellcolor{gray!25}5.86 & 5.29 & 19.24 & 16.44 & 14.85 & 12.79 \\
	    \midrule 
        E4 & {Small(18M+0.86M)} & {8} & \cmark & {SA-S2SND} & 5.90 & 5.01 & 16.33 & 13.68 & 12.93 & 10.84 \\
        \midrule
        E5 & {Medium(45.96M)} & {1} & \xmark & {S2SND} & 6.23 & 5.46 & 18.57 & 16.06 & 14.53 & 12.59 \\
        \midrule
        E6 & {Medium(45.96M)} & {1} & \xmark & {S2SND}~\tnote{$*$} & 6.84 & 5.59 & 17.41 & 14.13 & 13.94 & 11.33 \\
        \midrule
        E7 & {Medium(45.96M+0.86M)} & {1} & \cmark & {SA-S2SND}~\tnote{$*$} & 5.88 &\cellcolor{gray!25} 4.48 & \cellcolor{gray!25}16.33 & \cellcolor{gray!25}12.95 & \cellcolor{gray!25}12.92 & \cellcolor{gray!25}10.40 \\
        \bottomrule
	\end{tabular}
	\begin{tablenotes}
        \item [$*$]We use the compound dataset. The lowest online and offline DERs are highlighted by the gray background.
    \end{tablenotes}
	\end{threeparttable}
	\vspace{-0.5cm}
\end{table*}

\subsection{Network Configurations}

The network has five components: extractor, channel-attention, encoder, DOA model, and two decoders. We implement S2SND-Small and S2SND-Medium, both using a ResNet-34~\cite{he2016deep} extractor. The small version uses residual widths \{32,64,128,256\}, and the medium \{64,128,256,512\}. The channel-attention module is a 2-block Transformer~\cite{vaswani2017attention} with 512-dim 8-head attention and 1024-dim feed-forward layers. The encoder is a Conformer~\cite{gulati2020conformer} with a kernel size of 15, and both decoders follow the speaker-wise decoder of S2SND~\cite{cheng2024sequence}. And the two model variants share the same 4-block structure but differ in their hidden dimensions: S2SND-Small (256-dim, 8-head attention; 512-dim FFN) and S2SND-Medium (384-dim attention; 768-dim FFN), totaling 16.56M and 45.96M parameters, respectively. The single- and multi-channel versions differ only by the added channel-attention module. The DOA model (0.86M) is used only for estimation and not updated during training.

\vspace{-0.3cm}
\subsection{Training and Inference Details}

For training, the audio is segmented into 8s windows with 2s overlaps, which are then normalized and converted into 80-dimensional log-Mel filterbanks (25-ms frame length, 10-ms shift). The output resolution is 10 ms~\cite{cheng2024sequence}, with a maximum of $N=30$ speakers. To ensure permutation invariance, input embeddings and activities are randomly shuffled with reassigned labels. Data augmentation includes Musan noise~\cite{snyder2015musan} and RIR reverberation~\cite{ko2017study}. Optimization uses AdamW~\cite{loshchilov2017decoupled} with BCE and ArcFace losses~\cite{deng2019arcface} ($s=32,m=0.2$). Experiments run on two RTX-A6000 GPUs. 

For inference, the system applies block-wise sliding windows with left/chunk/right contexts. And the online latency is 0.8s (0.64s+0.16s). DOA features are fused at the encoder, and final buffers enable both online decoding and offline re-scoring.

\vspace{-0.2cm}
\section{Results}
\subsection{DOA Analysis}

It is worth noting that the original SRP-DNN was developed for the 12-channel LOCATA dataset~\cite{locata} and thus not directly applicable to the 8-channel circular array of AliMeeting. To address this, we retrain the CRNN component of SRP-DNN under the AliMeeting configuration. Fig.~\ref{fig:both} illustrates DOA estimation results on both simulated LibriSpeech mixtures and real AliMeeting recordings. The simulated data show negligible DOA errors, while real meeting scenarios exhibit clearly separated azimuth estimates across different speakers. In fact, meeting participants are typically seated, resulting in limited elevation variation, while azimuth provides clear and discriminative spatial cues (Fig. ~\ref{fig:both}(b)). Therefore, azimuth is used as the primary spatial feature. Notably, SA-S2SND is not limited to azimuth-only inputs, and elevation can be incorporated without architectural changes when needed.

Furthermore, SRP-DNN’s IDL-based detection tracks at most two speakers per frame, we believe it is reasonable in practice, since frames with more than two active speakers are rare—only 6\% in AliMeeting, 0.5\% in AISHELL-4 and 3\% in AMI (see Table~\ref{exp:ratio}), and thus have a negligible impact on overall diarization performance.

\subsection{Results on Alimeeting Dataset}

Table~\ref{exp:ali} shows SA-S2SND performance under different training conditions on Alimeeting, from which several conclusions can be drawn.

First, adding DOA consistently improves over baselines for both small and medium model: total DER drops by 4.2\% online (16.03$\rightarrow$15.35) and 7.4\% offline (13.59$\rightarrow$12.59), with larger gains in multi-speaker cases, showing better robustness under complex conversational conditions. Second, E2 vs.\ E3 indicates DOA-based decoupling is more effective than channel fusion offline. Third, E3 vs.\ E4 shows further DER reductions (12.9\% online, 15.2\% offline), highlighting complementarity: channel-attention captures correlations, while DOA provides explicit spatial cues. As a result, E4 achieves the best performance among small models, with 19.3\%/20.3\% relative gains over E1 (online/offline). Finally, scaling to the medium model further improves performance (E5 vs. E1), and training with the compound dataset and DOA achieves the overall best results, demonstrating the scalability and effectiveness of the proposed SA-S2SND framework.

\vspace{-0.2cm}
\begin{table}[t]
	\centering
	\setlength{\tabcolsep}{12pt}
    \vspace{-0.2cm}
	\caption{Comparisons of our methods with others on the Alimeeting test sets (offline).}
	%\vspace{-0.2cm}
	\label{exp:s2snd_comp}
	\begin{threeparttable}[b]
	\begin{tabular}{lrr}
		\toprule
		\textbf{Methods} &  & \textbf{DER (\%)} \\	    
		\midrule
		\textbf{Offline} \\
		\quad Diaper~\cite{landini2024diaper} &  & 20.70 \\
		\quad EEND-EDA ~\cite{Broughton25} &  & 12.30 \\	
		\quad Pyannote.audio ~\cite{plaquet2023powerset} &  & 15.20 \\
		\quad EEND-M2F ~\cite{harkonen2024eend} & & 13.20 \\
		\quad EEND-TA + FT ~\cite{Broughton25} &  & 11.41 \\
        \quad WavLM-Large ~\cite{han2025efficient}~\tnote{$\dag$} &  & 10.80 \\
	    % ---------------- % 
        \bottomrule
	    \quad S2SND (E6 in Table~\ref{exp:ali}) &  &  11.33 \\
        \quad SA-S2SND (E7 in Table~\ref{exp:ali}) &  &  \textbf{10.40} \\	
	    % ---------------- % 
		\bottomrule
	\end{tabular}
    	\begin{tablenotes}
        \item [$\dag$]State-of-the-art by submission. 
    \end{tablenotes}
	\end{threeparttable}

	\vspace{-0.5cm}
\end{table}

\subsection{Comparison with Other Existing Methods}
Table ~\ref{exp:s2snd_comp} compares our proposed methods with the previous results on the Alimeeting dataset. In the offline
scenario, our proposed SA-S2SND obtain the lowest DERs of
10.40\%. The WavLM-large method employs a pretrained WavLM with 63.3M parameters and uses a 16s window for both training and inference (DER=15.1\% for 8s window). Despite using a shorter 8s window and without pretraining models, our method achieves state-of-the-art performance.

\section{Conclusions}
\label{sec:con}

This work proposes SA-S2SND, a novel spatially-augmented diarization system that integrates direction-of-arrival (DOA) cues into a sequence-to-sequence neural diarization. By incorporating DOA features and a staged training strategy, the model unifies both single- and multi-channel inputs, supporting both online and offline inference. Evaluated on the AliMeeting test set, SA-S2SND achieves significant reductions, particularly when enhanced with channel attention. Future work will focus on improving multi-speaker DOA robustness and further enhancing overall system performance.

\section{Generative AI Use Disclosure}
Generative AI tools were used for limited language editing purposes, including improving clarity and correcting grammatical issues. No substantive content, analysis, or conclusions were generated by AI. The authors remain fully responsible for the content of this manuscript.

\begin{small}
\bibliographystyle{IEEEtran}
\bibliography{refs}   
\end{small}

\end{document}